\documentstyle[12pt]{article}
\begin{document}
%\begin{flushright}
%ROME prep. 94/1000 \\
%\end{flushright}
\title{Gribov Noise on the Lattice Axial Current Renormalisation
Constant}
\author{Maria Luigia Paciello$^{(1)}$,Silvano Petrarca$^{(1,2)}$,\\
        Bruno Taglienti$^{(1)}$ and Anastassios Vladikas$^{(3)}$
\\[1.5em]
$^{(1)}$INFN, Sezione di Roma {\it La Sapienza}\\
P.le A. Moro 2, 00185 Roma, Italy\\[0.5em]
$^{(2)}$Dipartimento di Fisica,\\
Universit\`a di Roma {\it La Sapienza},\\
P.le A. Moro 2, 00185 Roma, Italy\\[0.5em]
$^{(3)}$INFN, Sezione di Roma II,\\
II Universit\`a di Roma {\it Tor Vergata},\\
Via della Ricerca Scientifica 1, 00133 Roma, Italy
}

\maketitle
\newpage

\begin{abstract}
We study the influence of Gribov copies, in the Landau
gauge, on the lattice renormalisation constant $Z_A$ of the axial current,
obtained from a Ward identity on quark state correlation functions,
with the Clover action, in quenched $SU(3)$ gauge theory.
A comparison between the gauge invariant determination of $Z_A$ and
the gauge dependent one is discussed. On a $16^3 \times 32$ lattice
at $\beta = 6.0$ and with $K = 0.1425$,
the values, on a sample of 36 configurations, are: $Z_A$ = 1.08(5)
(gauge dependent calculation) and $Z_A$ = 1.06(2) (gauge independent
calculation).
We find that the residual gauge freedom associated to Gribov copies
induces observable effects, which, at the level of numerical
accuracy  of our simulation, are included in the statistical
uncertainty inherent in a Monte Carlo simulation.
Doubling the statistics suggests that the fluctuation due to the
lattice Gribov ambiguity scales down at least as fast as a pure
statistical error.
\end{abstract}
\vfill
\newpage

\section{Introduction}

In continuum non Abelian field theories, most popular choices
of fixing the gauge (e.g. Landau, Coulomb) suffer from the Gribov
ambiguity \cite{Gribov}. It is now well established that this
problem also affects the lattice formulation of these
theories \cite{mari}-\cite{Deforcra}. This problem has been
neglected for a long time because, in principle, the computation
of gauge invariant operators in compact lattice theories does not
require gauge fixing. Fixing the gauge is, however, necessary in
several cases. Monopole studies in SU(2) pure gauge
theory have been done in the unitary gauge and the effect
of the Gribov ambiguity on the number of SU(2) monopoles
has been investigated \cite{amst,jap}. The authors conclude that,
in their case,
the Gribov noise does not exceed the statistical uncertainty.
In SU(3) gauge theory, gauge fixing is essential in the
the computation of gauge dependent quantities, such as gluon and
quark propagators. There are now several studies of lattice propagators.
The gluon propagator has been calculated in \cite{mand_g}-\cite{bern_q}
with the aim of studying the mechanism through which the gluon may
become massive at long distances.
More recent attempts have investigated its behaviour as a function
of momentum \cite{mar_g,bern_g}. Analogous studies have also been
performed on the quark propagator (see, for example \cite{bern_q}).
In practice, there are also cases in which it
is convenient to implement a gauge dependent procedure for the computation
of gauge invariant quantities \cite{Parisi}-\cite{Shar}. For example,
smeared fermionic interpolating operators are widely being used in lattice
QCD spectroscopy and phenomenology, in order to optimise the
overlap of the lower-lying physical state with the operator. The
point-splitted smeared operators are gauge dependent, and therefore the gauge
must be fixed before they are calculated.
In particular, the calculation of
the decay constant of the $B$ meson in the static approximation, in which the
$b$-quark has infinite mass, requires the computation of
the two point correlation function of the axial current.
The isolation of the lightest state at large times is not possible if local
(gauge invariant) operators are used.
A nice way out consists in smearing the bilocal operator over a small
cube and extracting $f_B$ by forming suitable ratios of smeared and local
correlation functions \cite{Allton}. This is an explicitly gauge dependent
procedure which is most naturally carried out in the Coulomb gauge. In
ref.\cite{usgrib2} the smeared - smeared correlation functions on a few
individual configurations were computed. Two Gribov copies were
produced per configuration. The Gribov noise on individual configurations
was found to vary from $1\%$ to $50\%$ depending on the time-slice,
which implies that
it may still be a considerable effect after averaging over configurations.
However, it was not possible to estimate its effect beyond individual
configurations. The reason is that in such a study other sources of
error dominate, such as the systematic error arising from fitting
the exponential decay of the correlation function with time. Thus the
isolation of the Gribov noise is difficult.
\footnote{More recently, a high statistics
study of $f_B$ in the static limit \cite{apefb} uses a different
method for constructing ratios of smeared and local correlators which
avoids fitting. This method, however, requires a large temporal extention
of the lattice.}
\par
In this paper we study a different physical quantity, namely the
renormalisation constant $Z_A$ of the lattice axial current.
A knowledge of these renormalisation constants is necessary for matching
the matrix elements computed using lattice simulations to those required
in a definite continuum renormalisation scheme. Provided that the lattice
spacing
is sufficiently small it is possible to calculate these renormalisation
constants in perturbation theory. For a more reliable determination
of these constants it has been suggested to impose
the chiral Ward identities of $QCD$ non-perturbatively \cite{ks,boc}.
Here we focus our attention on the determination of
the r${\hat{\rm{o}}}$le of the Gribov ambiguity in the
calculation of $Z_A$, obtained from quark state correlation functions.
A recently proposed method to determine $Z_A$ and other
renormalisation constants, based on truncated quark Green functions
in momentum space \cite{dallas} can also in principle be afflicted
by Gribov fluctuations.
\par
Since reasonably small errors are expected, in this kind of calculations,
it is crucial to investigate the r$\hat{\rm{o}}$le of the Gribov noise.
Moreover, the renormalisation constant
$Z_A$ of the axial current is particularly well suited to the study
of the Gribov fluctuations, mainly for
two reasons. Firstly, $Z_A$ can be obtained
from chiral Ward identities in two distinct ways:
a gauge independent one, which consists in taking the matrix elements
between hadronic states, and a gauge dependent one, which consists
in taking the matrix elements between quark states. Hence,
there is an explicitly gauge invariant estimate of $Z_A$ which is
free of Gribov noise and which can be directly compared to the gauge
dependent, Gribov affected, estimate. The second advantage is that $Z_A$ is
obtained by solving a first degree algebraic equation for each lattice time
slice, thus avoiding the usual systematic errors arising from fitting
exponentially decaying signals in time.

\section{Chiral Ward identities for $Z_A$}

The theoretical framework for the non-perturbative evaluation of
$Z_A$ for Wilson fermions, has been developed in \cite{boc}.
The renormalisation constant is obtained through Ward
identities generated by axial transformations. A first application
of these techniques in numerical simulations using the Wilson action
was attempted in \cite{maiani}.
The extension of these methods to the $O(a)$ improved Clover action
\cite{SW} ($a$ is the lattice spacing) was presented in \cite{wiza},
which we follow most closely. Here we only give
a brief outline of the results which are essential to our work.
\par
In this study terms that, close to the continuum limit, are effectively of
$O(a)$ are eliminated by using the Clover action \cite{Heatlie} and rotating
all quark fields of the matrix elements
according to the "improved improvement" prescription of \cite{impimp}:
\begin{eqnarray}\psi
\rightarrow (1- \frac{ra}{2}\gamma\cdot {\stackrel{\rightarrow}{D}})\psi
\,\,\,\,\,\, ; \,\,\,\,\,\,
\bar{\psi} \rightarrow \bar{\psi} (1+\frac{ra}{2}\gamma\cdot
{\stackrel{\leftarrow}{D}})
\label{rot}
\end{eqnarray}
($r$ is the Wilson  term parameter; in this work $r=1$).
The two fermion local operators considered in the following
are the axial and vector
currents and the pseudoscalar density:
\begin{eqnarray}
A_\mu^f(x) \equiv \bar\psi(x)\gamma_\mu\gamma_5\frac{\lambda^f}{2}\psi(x)
\nonumber \\
V_\mu^f(x) \equiv \bar\psi(x)\gamma_\mu\frac{\lambda^f}{2}\psi(x)
\nonumber \\
P_5^f(x) \equiv \bar\psi(x)\gamma_5\frac{\lambda^f}{2}\psi (x)
\end{eqnarray}
($f$ is a flavour label and the notation is generic for any quark fields
$\psi$ and $\bar\psi$).
In order to ensure that the lattice axial current $A_\mu^f$ has
the correct chiral properties, it is normalised by a renormalisation constant
$Z_A$ \cite{boc}; this implies that
\begin{equation}
2\rho = \frac{\partial_4\int d^3{\vec y}<\ A_4^{f}(\vec y, t_y)
P_5^{\dagger\,f}(\vec 0,0)\ >}
{\int d^3{\vec y}<\ P_5^{f}(\vec y, t_y) P_5^{\dagger\,f}(\vec 0,0)\ >}
= \frac{2m}{Z_A}
\label{eq:twomzal}
\end{equation}
(m is the bare quark mass and $|P>$ the pseudoscalar state).
As can be seen from the above equation, $\rho$ is gauge invariant.
\par
The gauge dependent calculation of $Z_A$ relies on the following
Ward identity for quark Green functions
\cite{boc,wiza}
\begin{eqnarray}
2\rho Tr \left[ \int d^4x \int d^3\vec y
<\ u_\alpha (y)\,\left(\bar u(x)\gamma_5 d(x)\right)
\,\bar d_\beta (0)\ > \right] \nonumber \\
= \left(
\frac{1}{Z_{A}}-{\rho ra}\right) Tr \left[ \int d^3\vec y
< (\gamma_5 d(y)\bar d(0)\, +\, u(y)\bar u (0)\gamma _5) >_{\alpha,\beta}
\right]
\label{eq:wiqp}
\end{eqnarray}
In eq.(\ref{eq:wiqp}) we work explicitly with up and
down quark fields with spinor labels $\alpha$ and $\beta$.
The trace is over colour indices.
The expectation values on both sides of (\ref{eq:wiqp}) are evaluated
as functions of $t_y$.
Taking the value of $2\rho$ obtained using
eq.(\ref{eq:twomzal}), $Z_{A}$ can then be determined.
In order to enhance the signal, we add in both sides
of (\ref{eq:wiqp}) the four contributions $(\alpha,\beta)$=(1,3),
(3,1), (2,4) and (4,2), which were found to give the clearest signal
\cite{wiza}. A plateau in $t_y$ is typically obtained and $Z_A$ is estimated
from it. The crucial point is that
both sides of eq.(\ref{eq:wiqp}) are gauge dependent, and thus this
determination of $Z_A$ is in principle sensitive to the Gribov noise.
\par
A gauge invariant determination of $Z_A$ is obtained through the
Ward identity
\begin{eqnarray}
\lefteqn{2\rho\int d^4x\int d^3\vec y
<\ P_5^{f}(x)A^{g}_\nu(y)V^{h}_\rho (0)\ >}\nonumber\\
& = & -i\left( \frac{Z_V}{Z_A ^2}-\rho ra \right) f^{fgl}
\int d^3\vec y<\ V^{l}_\nu(y)V^{h}_\rho (0)\ >\nonumber\\
& & -i\left( \frac{1}{Z_V}-\rho ra \right) f^{fhl}
\int d^3\vec y<\ A^{g}_\nu(y)A^{l}_\rho (0)\ >
\label{eq:wivaint}
\end{eqnarray}
In the above equation, the vector current renormalisation constant $Z_V$
is also needed. For the Clover action, $Z_V$ is calculated with the aid
of the so-called conserved and improved vector current
\begin{eqnarray}
V_\mu^{CI}(x) & = & \frac{1}{4}\left[ \bar\psi(x)(\gamma_\mu-r)
U_\mu(x)\psi(x+\hat\mu) +
\bar\psi(x+\hat\mu)(\gamma_\mu + r)
U^\dagger_\mu(x)\psi(x) \right] \nonumber\\
& &\ \ \ \ \ \ \ \ +(x\rightarrow x-\hat\mu)
+\frac{r}{2} \sum_{\rho}\partial_\rho\left(\bar\psi(x)\sigma_
{\rho\mu}\psi(x)\right)
\label{eq:vci}
\end{eqnarray}
Since the current is conserved, its renormalisation constant is
precisely 1, and, since it is ``improved", its matrix elements have no
corrections of $O(a)$. The normalisation constant $Z_V$ of the local
vector current $V_\mu (x)$ is determined through the ratio
of the vacuum to vector state
matrix elements of the two vector currents \cite{msv}:
\begin{eqnarray}
Z_V \equiv \frac{<0|V_\mu^{CI}(0)|V_\mu>}{<0|V_\mu(0)|V_\mu>}
\label{eq:zvrat}
\end{eqnarray}
(here $|V_\mu>$ denotes the vector state).
By calculating all correlation functions
of eq.(\ref{eq:wivaint}) and $Z_V$ from eq.(\ref{eq:zvrat})
and by requiring that eq.(\ref{eq:wivaint})
holds at each $t_y$, we can determine $Z_A$ in an explicitly gauge
invariant fashion for which the Gribov ambiguity is irrelevant.
\par
One more comment is in place here: the terms proportional
to $\rho r a$ on the right-hand-side of eqs.(\ref{eq:wiqp})
and (\ref{eq:wivaint})
arise from the rotations of the fermion fields defined in eq.(\ref{rot}),
which are inherent to Clover action improvement \cite{Heatlie}.
The $\gamma \cdot D$ rotations, combined with the equations of
motion, generate contact terms which, to $O(a)$, give rise to the
terms proportional to $\rho r a$ \cite{wiza}.

\section{Lattice Gauge Fixing and Gribov Copies}

Gauge fixing and the generation of Gribov copies on
the lattice is by now a standard procedure. Given a
thermalised configuration generated by a Monte Carlo simulation,
the Landau gauge is fixed by minimising the functional \cite{Davies}
\begin{equation}
F [U^g] \equiv - \frac{1}{V}
Re \ Tr  \sum_{\mu=1}^{4} \sum_{n} U_{\mu}^{g}(n)
\label{eq:bigf}
\end{equation}
with respect to $g$. Here
$V$ is the lattice volume and $U_{\mu}^{g}(n) \equiv g(n)
U_{\mu}(n) g(n + \mu)^{\dagger}$ is the compact
$SU(3)$ gauge field, gauge transformed by a local gauge transformation
$g(n)$. The extrema of $F$ correspond to configurations that satisfy the
gauge condition $\partial_\mu A^g_\mu = 0$
in discretised form. The minimisation
of $F$ is obtained through iteration: each lattice site is visited and $F$
is minimized locally. After several lattice sweeps $F$ becomes constant,
and the gauge is fixed. This is nothing
else but a discretised analogue of the continuum formulation \cite{Zwa},
according to which the local minima
of the functional ${\cal{F}} [A^g_\mu] = - \sum_{\mu=1}^4 Tr \int d^4x
(A^g_\mu)^2 $ with respect to $g(x)$ correspond to configurations in the
Landau gauge.
Even if there is a straightforward correspondence between the lattice
and continuum gauge fixing procedures, it is interesting to note that on
the lattice, because of discretisation, there can be more minima than in
the continuum \cite{Testa}.
\par
The production of Gribov copies
consists in generating gauge equivalent configurations,
by applying random gauge transformations to our thermalised
link configuration \cite{Davies}, \cite{mari}. The
gauge is then fixed and $F$ is measured. Since it is
a gauge dependent quantity, a different value of $F$ for two
gauge equivalent, gauge fixed configurations means that they are Gribov
copies. The over-relaxation technique of \cite{Mand2} (which consists
in accelerating the gauge fixing algorithm by raising the gauge
transformation $g(x)$ to a real tunable power $\omega$ at every
iteration) has been implemented at fixed $\omega$. The over-relaxation
itself can be used for the generation of Gribov copies, by varying the value
of $\omega$, as proposed in \cite{usgrib3}. In this work we have opted for
the random gauge transformation method.

\section{Results}

We work in the framework of the quenched approximation with the Clover
action of SU(3) gauge theory.
The lattice volume is $V = 16^3 \times 32$ and $\beta = 6.0$. After $3000$
thermalising sweeps, 36 configurations were generated, separated by $1000$
sweeps. An $8$ hit Metropolis algorithm was used.
For each thermalised configuration, we generated $6$ Gribov copies.
This was done by fixing the gauge both on the original configuration
and on $5$ gauge equivalent replicas, obtained by applying random
gauge transformations.
It is remarkable that, in our case, each random gauge transformation produced
a Gribov copy. This high probability to find Gribov copies is a
characteristic of large volume lattices in the confined region, as
discussed in \cite{usgrib3,mar_ros}.
We fix the Landau gauge, using the over-relaxation
algorithm suggested in \cite{Mand2} at fourth order in the over-relaxation
parameter $\omega = 1.72$.
The stopping condition we have imposed is that the relative variation
$\delta F / F$, of the minimised functional $F$, between two consecutive
gauge fixing sweeps be less than $10^{-8}$.
  This value guarantees a good quality of the gauge fixing, allows us to
distinguish Gribov copies, and it is typically reached after a number
of sweeps which varies between
500 and 1500. The gauge fixing was done on an IBM Risc
6000/550 equipped with 128 Mbyte of RAM memory and with a CPU working at
42 MHz; with this machine a single gauge fixing sweep takes about 40 s.
The quark propagators, rotated as indicated by eq.(\ref{rot})
were obtained at a Wilson hopping parameter value of $K=0.1425$, which,
for the Clover action, corresponds to a pion of roughly 900 MeV.
\par
Before passing to a detailed discussion of the Gribov
noise, it is appropriate to present a comparison of the results
for $Z_A$, obtained in a gauge invariant way (see eq.(\ref{eq:wivaint})),
to those based on the gauge dependent identity of eq.(\ref{eq:wiqp}).
This calculation has been already presented in \cite{wiza}, on the
first 18 configurations of our ensemble; here we have doubled the statistics.
Moreover, we have re-gauged the first 18 configurations in order to reach the
better quality of gauge fixing  required for this study.
In Fig.(1) we show the behaviour of the
two estimates of $Z_A$, as a function of $t_y$,
calculated on the same ensemble of 36 configurations.
The gauge invariant values of $Z_A$ show a flat behaviour
already at $t_y=5$ and with small error bars. The new value of $Z_A$,
obtained with the gauge independent technique,
over 36 configurations, is $Z_A=1.06 (2)$ to be compared with the old one
obtained over 18 configurations: $Z_A=1.09 (3)$ (see ref.\cite{wiza}).
\par
In the gauge dependent case, we have taken
the average over the Gribov copies, in the way
that will be discussed below. In this case,
the $Z_A$ behaviour becomes flat only at $t_y=10$
showing a large sensitivity to the contact terms of the Ward identity
at small $t_y$ values. The new value of $Z_A$,
obtained from the gauge dependent Ward identity, as can be seen from
Fig.(1) is $Z_A = 1.08(5)$; to be compared to the value
obtained from 18 configurations, $Z_A = 1.14(8)$ (see ref.\cite{wiza}).
As already stressed in \cite{wiza},
this method gives results that are compatible within the errors
with the gauge independent ones. Moreover,
the error of the gauge invariant calculation
of $Z_A$ is always smaller than the error of its gauge dependent
counterpart. This is due to the fact that the quark state
correlation functions fluctuate more than the gauge invariant correlation
functions, but it may also indicate the presence of another
effect, which is probably the Gribov ambiguity.
In the hypothetical case of two determinations of $Z_A$, affected by the
same statistical error, the gauge dependent estimate should fluctuate more
due to the Gribov noise. Then the amount of Gribov noise could be estimated
as the difference (in quadrature) of the two errors.
\par
Normally, in a standard simulation of gauge dependent
quantities, one does not generate Gribov copies. Consequently,
one measures a given quantity by taking the average and error
over the ensemble of the gauge fixed configurations that have been
generated. This implies a particular and arbitrary
choice of Gribov copies. The error estimated, for example, by a jacknife
method, is not purely statistical as it
implicitly contains the uncertainty due to the
particular choice of a Gribov copy.
\par
In our case, having generated $G=6$ copies
for each of the
$N=36$ thermalised configurations, there are $G^N$ possible combinations
that we may consider when forming a particular ensemble. To the best of
our knowledge, the
distribution of the Gribov copies of a given configuration is unknown;
thus the weight to be associated to it is arbitrary.
Moreover, any technique used to generate different Gribov copies selects
a particular copy in a completely uncontrolled way. Hence, we assume
that the particular
choice of different combinations of Gribov copies when forming a
statistical ensemble is arbitrary. In order to exhibit the effect
of such arbitrariness, we show in Fig.(2)
the behaviour of $Z_A(t_y)$ for $4$ arbitrary choices of copies.
The $4$ different behaviours are compatible,
and the same is true for the jacknife errors. It is
clear, however, that the presence of Gribov copies is a visible effect;
each of the six estimates of $Z_A$ shown has a slightly
different profile as a function of $t_y$.
\par
We now expose the procedure we implemented for taking into account
the Gribov ambiguity. The gauge dependent traces of the
two and three point correlation
functions appearing in eq.(\ref{eq:wiqp}), calculated on a single
configuration $c$ and for a particular Gribov copy $g$ are denoted by:
\begin{eqnarray}
T_2(t_y ; c,g) = Tr \left[ \int d^3\vec y
(\gamma_5 d(y)\bar d(0)\, +\, u(y)\bar u (0)\gamma _5 )
\right] \nonumber \\
T_3(t_y ; c,g) = Tr \left[ \int d^4x \int d^3\vec y
\, u(y)\,\left(\bar u(x)\gamma_5 d(x)\right) \,\bar d(0)\ \right]
\label{eq:t2t3a}
\end{eqnarray}
(in the above equations the Dirac indices have been implicitly
averaged over, as explained in Sect.(2)).
Then, for a given configuration $c$, we consider our "best estimate"
of these matrix elements to be their average over the $G=6$ Gribov copies:
\begin{eqnarray}
\bar{T}_{2,3}(t_y ; c) = \frac{1}{G} \sum_{g=1}^{G} T_{2,3}(t_y;c,g)
\label{eq:t2t3g}
\end{eqnarray}
The average of the above $\bar{T}$'s over the $36$ configurations
will be taken as our "best estimate" $<\bar{T}>$ for the gauge dependent
traces. On the other hand, $\rho$, being gauge invariant,
does not depend on a particular
choice of Gribov copies, but only on the configuration ensemble.
Then $Z_A(t_y)$ is obtained by applying eq.(\ref{eq:twomzal})
as follows
\begin{eqnarray}
Z_A (t_y)^{-1} = 2 \rho(t_y)
\frac{<\bar{T}_3(t_y)>}{<\bar{T}_2(t_y)>} + \rho(t_y) r a
\label{eq:zam1}
\end{eqnarray}
The error is obtained by a standard jacknife method performed on
the quantities $<\bar{T}_2(t_y)>$,$<\bar{T}_3(t_y)>$ and $\rho(t_y)$,
by decimating one configuration at a time.
This is the gauge dependent $Z_A(t_y)$ result shown in
Fig.(1).
\par
We want to stress that the values of $<\bar{T}_2>$ and $<\bar{T}_3>$
fluctuate more than their ratio $<\bar{T}_3> / <\bar{T}_2>$.
The latter quantity is gauge invariant, according to eq.(\ref{eq:zam1}).
For example, at $t_y = 10$, $\delta\bar{T}_2 / <\bar{T}_2> = 5.6\%$,
$\delta\bar{T}_3 / <\bar{T}_3> = 5.1\%$ and
$\frac{\delta(\bar{T}_3 / \bar{T}_2)}{<\bar{T}_3> / <\bar{T}_2>} = 2.3\%$.
The strong reduction of the relative error indicates
a great sensitivity of $\bar{T}_{2,3}$ to the Gribov
fluctuation, as opposed to a relative stability of the gauge invariant
quantities.
\par
In order to estimate the uncertainty arising from a particular choice
of copies, out of the $6^{36}$ possible ones,
we have applied a procedure which takes this arbitrariness into account.
We have chosen randomly $10^4$ combinations of copies of our
ensemble and have calculated $Z_A(t_y)$ for each one of these combinations
at fixed $t_y$.
 The histogram of the values obtained for $Z_A(t_y)$
is well fitted by a Gaussian, the r.m.s. width of which is taken as
an estimate of the fluctuation.
In Fig.(3 a) we compare the jacknife error of our "best
estimate" to the width of the Gaussian. We see that for all $t_y$
the width is smaller than the jacknife error. This implies that
the fluctuations induced by
the particular choice of Gribov copy when forming the ensemble
are small and do not overcome the statistical uncertainty.
\par
As it is also important to understand how the above behaviour scales with
increasing the number of configurations, we have performed the same
analysis for the first $18$ configurations (half of our ensemble).
The result is shown in Fig.(3 b). Comparing
Fig.(3 a) to Fig.(3 b), we note that both errors scale at least as
$\sqrt{2}$. Thus, within our moderate statistics, we find that the error,
even if affected by the Gribov ambiguity, decreases with increasing
configuration number.
\par
In conclusion, our investigation, even within its limitations, shows that,
on the lattice, the residual gauge dependence associated with the Gribov copies
does not generate an overwhelming fluactuation of the $Z_A$ measurements
performed by the gauge dependent method. Nevertheless the arbitrariness
associated to a particular choice of Gribov copies is a visible
effect which manifests itself, especially in the behaviour of $Z_A$
as a function of $t_y$. The jacknife error is greater in the gauge
dependent determination than in the gauge independent one. However,
in the former case, even though the error is afflicted by the Gribov
uncertainty, it is still decreasing with increasing statistics.
This implies that the uncertainty arising from the
Gribov ambiguity may be a secondary effect.
Analogous studies on different physical quantities and
renormalisation constants could further support this belief.

\section{Acknowledgements}

We warmly thank G.Martinelli, C.T.Sachrajda and M. Testa
for many discussions and partecipation throughout this work.
Useful discussions with
V.N. Gribov and A.J. van der Sijs are also greatfully acknowledged.
We also thank the IBM - Semea for providing us with 32 Mbytes of
memory.

\section{Figure Captions}

\par
FIGURE 1: Comparison of the gauge independent calculation of $Z_A(t_y)$
(diamonds) to the gauge dependent one (crosses). The errors are jacknife.
The crosses have been slightly displaced to help the eye. \\
FIGURE 2: The gauge dependent calculation of $Z_A(t_y)$ for 4
arbitrarily chosen Gribov copies. The errors are jacknife. \\
FIGURE 3: Comparison of the jacknife error (crosses) to the r.m.s.
Gaussian width (diamonds) due to the Gribov ambiguity (see text).
The crosses have been slightly displaced to help the eye.
Case (a) is with 36 configurations; case (b) with the first 18
configurations

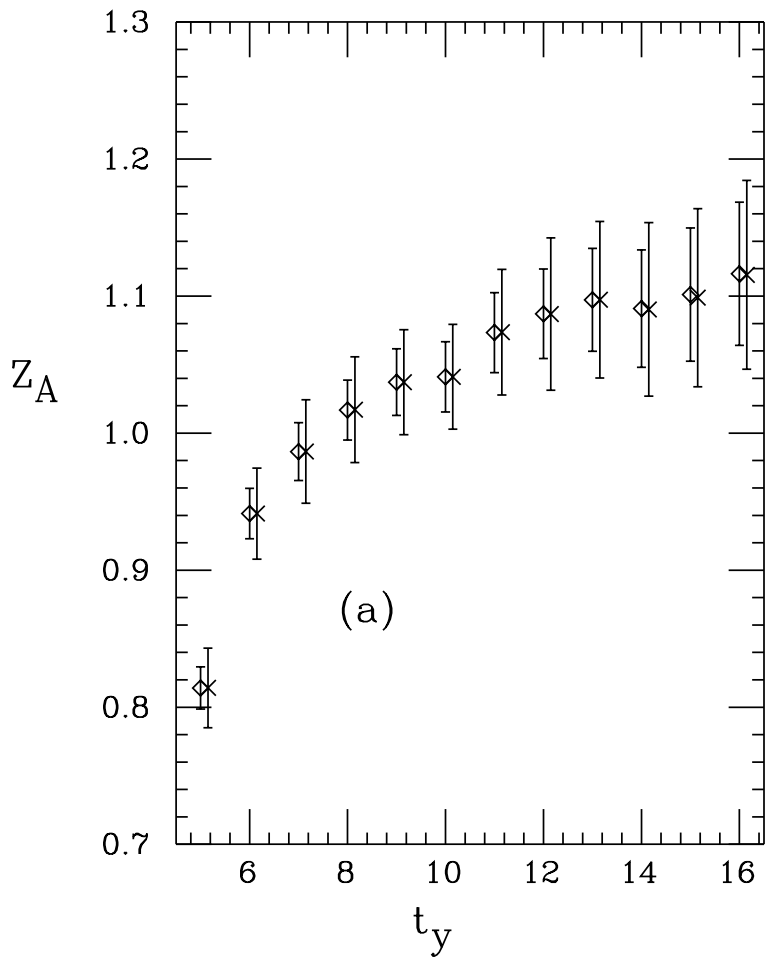
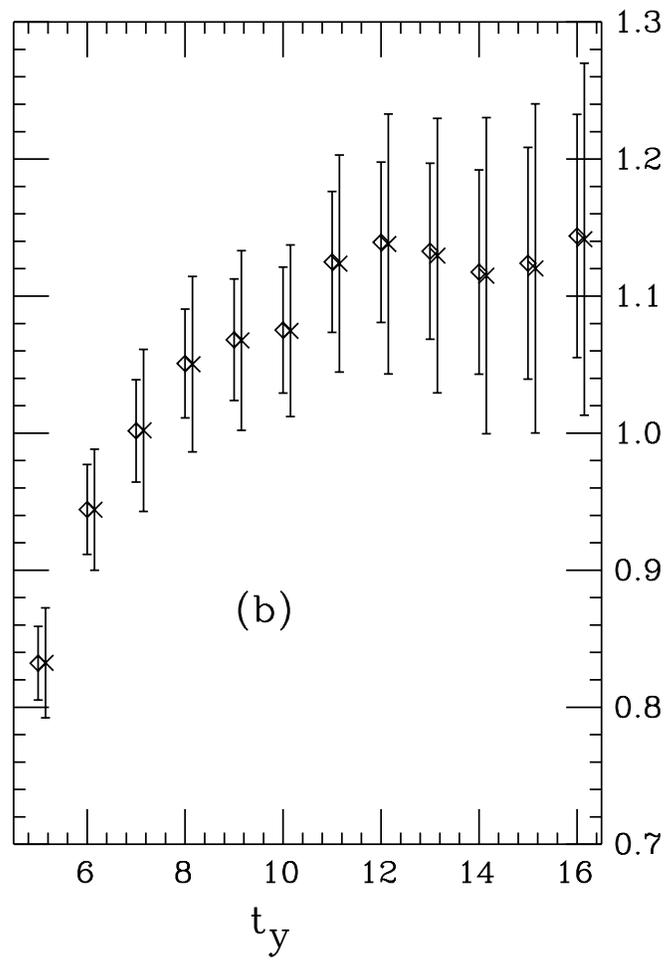
\end{document}